\documentclass[aps,floatfix,twocolumn,prl]{revtex4}
\usepackage{graphicx}
\usepackage{amsmath}
\usepackage{amssymb}

\begin{document}
\title{Bose-Einstein Condensates in Time Dependent Traps}
\author
{Y. Castin and R. Dum}
\address{
Ecole Normale Sup\'erieure, Laboratoire Kastler Brossel, 24, Rue Lhomond, F-75231 Paris Cedex 05, France\thanks{Laboratoire Kastler Brossel
is an unit\'e de recherche de l'Ecole Normale Sup\'erieure et de l'Universit\'e
Pierre et Marie Curie, associ\'ee au CNRS.}
}
\date{\today}

\begin{abstract}
We present analytical results for the macroscopic
wave function of a Bose-Einstein condensate 
in a time dependent harmonic potential. The evolution of the spatial density
is a dilatation, characterized by three scaling factors which allow a 
classical interpretation of the dynamics.
This approach is an efficient tool for
the analysis of recent experimental results 
on the expansion and collective excitation of a condensate.
\end{abstract}
\pacs{03.75.Fi,05.30.Jp}

\maketitle 

Recently the combination of laser cooling and evaporative cooling led
to the observation of Bose-Einstein condensation in dilute atomic vapors
\cite{BEC1,BEC3,BEC4}. The favored observation technique has been
a time of flight measurement: the trapping potential is rapidly
switched off and the spatial distribution of the expanding cloud is monitored.
In more recent experiments the condensates were collectively excited by a time
modulation of the trapping potential \cite{JILA,MIT}.
In these experiments the state of the condensate is strongly influenced by atomic
interactions, which must therefore be included in a theoretical treatment.
Work up to now consisted in the
numerical solution of the time dependent 
non-linear Schr\"odinger equation for the macroscopic
wave function of the condensate\cite{NUM}.
We present here analytical results which
allow a more lucid description of the condensate dynamics and an immediate
comparison with experiment.
To this end we introduce a quantum scaling transform \cite{SHLAP}
which is inspired by a model of a classical gas. 
Applying our results to time of flight measurements 
of expanding condensates \cite{BEC4} we obtain 
the scattering length of sodium. For condensates collectively
excited by a time modulation of the trapping potential we present an
{\it ab initio} calculation of the observed signal.

For dilute gases at low temperatures the
atomic interactions can be modeled by
a pseudopotential $g \delta(\vec{r}) $,
where $g>0$ is related to the $s$-wave scattering length 
$a$ by $g=4 \pi \hbar^2 a /m$ \cite{HUANG}.
We describe the trap by an anisotropic 
time dependent harmonic potential
\begin{equation} \label{harm}
U(\vec{r},t) ={1 \over 2} \sum_{j=1,2,3} {m \omega_j^2(t) r_j^2 }.
\end{equation}
We restrict the discussion to the case of zero temperature, which is a realistic
assumption for the experiments in \cite{BEC1,BEC3}.
The state of the condensate for a static trap 
can thus be described using a Hartree-Fock Ansatz: 
\begin{equation} \label{HARTREE}
{\left | {\Psi} {\,} \right \rangle}= {\left | {\Phi} {\,} \right \rangle} \otimes  \cdots \otimes {\left | {\Phi} {\,} \right \rangle} .
\end{equation}
The minimization of mean energy gives the time independent Gross-Pitaevskii equation
for $|\Phi\rangle$:
\begin{equation} \label{GPE}
\mu \Phi(\vec{r}) 
= (-{\hbar^2 \over 2m} \Delta + U(\vec{r},0)
					 + N g |\Phi(\vec{r})|^2)\Phi(\vec{r}).
\end{equation}
with $N-1\simeq N$.

In the regime where the atomic interactions are dominant
($Ng |\Phi(\vec{0})|^2\simeq\mu \gg \hbar\omega_j$ for $j=1,2,3$) 
we can use the Thomas-Fermi
approximation to solve (\ref{GPE}) \cite{BAYM}, that is we can
neglect the kinetic
energy term as compared to the interaction energy term. The result is
\begin{equation} \label{TFD}
\Phi(\vec{r}) \simeq \Phi_{TF}(\vec{r})=
\left({\mu-U(\vec{r},0)\over Ng}\right)^{1/2}
\end{equation}
when $\mu\geq U(\vec{r},0)$, and $\Phi(\vec{r})=0$ otherwise.
The chemical potential $\mu$
is determined by the normalization of ${\left | {\Phi} {\,} \right \rangle}$:
\begin{equation}
\label{mu}
\mu={1\over 2}\hbar\bar{\omega}\left(15Na\sqrt{m\bar{\omega}\over \hbar}
\right)^{2/5}
\end{equation}
where $\bar{\omega}=(\omega_1(0)\omega_2(0)\omega_3(0))^{1/3}$.

One can generalize the Hartree-Fock Ansatz (\ref{HARTREE})
to the time dependent case, in which $\Phi$ 
is a function of $t$.
A time dependent variational calculus leads to an 
(explicitly) time dependent Gross-Pitaevskii equation \cite{PINES,YCRD}:
\begin{equation} \label{GPET}
i \hbar \partial_t \Phi(\vec{r},t) 
= (-{\hbar^2 \over 2m} \Delta + U(\vec{r},t)
					 + N g |\Phi(\vec{r},t)|^2)\Phi(\vec{r},t).
\end{equation} 
In the solution of (\ref{GPET}) a Thomas-Fermi type approximation 
is not  directly applicable, because the time variation
of the trapping potential would convert potential energy
into kinetic energy, which therefore could
no longer be neglected.
In this paper we
identify a unitary transform which eliminates the extra kinetic energy.

We first introduce a model of a classical gas 
in which each particle experiences a force 
\begin{eqnarray}
\vec{F}(\vec{r},t)= -\nabla (U(\vec{r},t)+g \rho_{cl}(\vec{r},t)) 
\end{eqnarray}
where $\rho_{cl}(\vec{r},t)$ is the spatial 
density in the gas normalized to $N$.
At $t=0$ the equilibrium condition $\vec{F} = 0$ gives
$\rho_{cl}(\vec{r},0)=N |\Phi_{TF}(\vec{r},0)|^2$, that is the classical
solution for the steady state density
coincides with the quantum solution in the Thomas-Fermi limit.
For $t>0$ the exact solution for the classical model
can be obtained for the class of potentials (\ref{harm}); in this
case, the gas merely experiences a dilatation, 
any infinitesimally small fraction of the expanding cloud moving 
along a trajectory  
\begin{equation}
\label{TRAJ}
R_j(t)=\lambda_j(t) R_j(0)\ (j=1,2,3). 
\end{equation}
From this Ansatz we obtain for the evolution of the spatial density:
\begin{equation} \label{dilat}
\rho_{cl}(\vec{r},t) = 
{1 \over \lambda_1(t) 
\lambda_2(t) \lambda_3(t)}\rho_{cl}(\{r_j/\lambda_j(t)\}_{j=1,2,3},0).
\end{equation}
Newton's law $m \ddot{R_j}(t) =F_j(\vec{R}(t),t)$ applied
for the trajectory (\ref{TRAJ}) implies:
\begin{multline}
\label{NEWTON}
m \ddot{\lambda_j}(t) R_j(0)=-(\partial_{r_j}U)(\vec{R}(t),t) \\
+{1\over \lambda_j\lambda_1\lambda_2\lambda_3} 
(\partial_{r_j}U)(\vec{R}(0),0) \ \ (j=1,2,3).
\end{multline}
From Eq.(\ref{dilat}) we have expressed the gradient of $g\rho_{cl}(t)$ 
in terms of $\nabla g\rho_{cl}(t=0)=-\nabla U(t=0)$.
For the harmonic potentials $U$ of Eq.(\ref{harm}) both sides of Eq.(\ref{NEWTON})
are proportional to $R_j(0)$. Eq.(\ref{NEWTON}) therefore holds for any $\vec{R}(0)$
and the Ansatz (\ref{TRAJ}) is self-consistent provided that
the scaling factors $\lambda_j(t)$ satisfy:
\begin{eqnarray} \label{lambda}
\ddot{\lambda_j} &=& {\omega_j^2(0) \over \lambda_j 
\lambda_1\lambda_2\lambda_3} 
	- \omega_j^2(t) \lambda_j \ \ \ \ \ (j=1,2,3).
\end{eqnarray}
The initial conditions are $\lambda_j(0) =1$ and since
the gas is initially at rest, $\dot{\lambda}_j(0)=0$.
Taking the time derivative of Eq.(\ref{TRAJ}) and eliminating the initial
position by Eq.(\ref{TRAJ}) we obtain for
the local velocity of the expanding cloud:
\begin{equation} \label{velo}
v_j(\vec{r},t) = r_j {\dot{\lambda_j}(t) \over \lambda_j(t)}.
\end{equation}
The equations (\ref{lambda},\ref{velo}) do not depend on the interaction strength
$g$. The $g$ dependence is entirely contained in the initial spatial
density of the gas \cite{HYDRO}.

This classical solution motivates the Ansatz for the solution 
of the quantum equation (\ref{GPET}):
\begin{eqnarray}
\Phi(\vec{r},t)=&& 
    e^{-i\beta(t)} e^{i m\sum_j r_j^2 \dot{\lambda}_j(t) /
			(2\hbar \lambda_j(t))} \times \nonumber \\
&&	{\tilde{\Phi}(\{r_k/\lambda_k(t)\}_{k=1,2,3},t) \over 
    \sqrt{\lambda_1\lambda_2\lambda_3}} \label{Ansatz}
\end{eqnarray}
Equation (\ref{Ansatz}) is a unitary transform combining a scaling in
$\vec{r}$ and a gauge transform.
The gauge transform subtracts from the momentum operator
$\hat{\vec{p}}$ the local momentum of the expanding classical
gas (\ref{velo}):
\begin{equation}
\hat{p_j} \rightarrow \hat{p_j} + m {\dot{\lambda_j}(t) \over \lambda_j(t) }
\hat{r_j}.
\end{equation}
The scaling transform mimics the dilatation (\ref{dilat})
obtained in the classical case.
We insert the Ansatz (\ref{Ansatz}) in Eq.(\ref{GPET}). For the convenient choice
of the global phase factor $e^{-i\beta(t)}$, 
$\hbar \dot{\beta}=\mu /\lambda_1(t)\lambda_2(t)\lambda_3(t)$,
we obtain after some algebra the following 
time evolution for $\tilde{\Phi}(\vec{r},t)$:
\begin{multline} \label{GPETILDE}
\left[ i \hbar \partial_t + {\hbar^2 \over 2m}
\sum_j {1 \over \lambda_j^2(t)} \partial^2_{r_j}\right]
\tilde{\Phi}(\vec{r},t)= \\
{ 1 \over \lambda_1(t)\lambda_2(t)\lambda_3(t)} 
\left[-\mu+ U(\vec{r},0)+ N g |\tilde{\Phi}(\vec{r},t)|^2\right] 
\tilde{\Phi}(\vec{r},t)
\end{multline}
with the initial condition $\tilde{\Phi}(\vec{r},0)=\Phi(\vec{r},0)$.
In the Thomas-Fermi regime the right hand side 
of Eq.(\ref{GPETILDE}) is initially very small; the
kinetic energy terms on the left hand side are also small initially,
and are expected to remain small in time, 
since the extra kinetic energy due to a 
change in the trapping potential has been absorbed in the unitary transform
(\ref{Ansatz}). We therefore expect that $\tilde{\Phi}(\vec{r},t)$ 
evolves weakly in time.

To show this point more rigorously we split $\tilde{\Phi}(\vec{r},t)$
into $\Phi(\vec{r},0)
+\delta\tilde{\Phi}(\vec{r},t)$. From Eq.(\ref{GPETILDE}),(\ref{GPE})
we find that 
$\delta\tilde{\Phi}(\vec{r},t)$ obeys a non-linear inhomogeneous equation 
with a source term given by:
\begin{equation}
S(\vec{r},t)= -{\hbar^2\over 2m}\sum_{j=1}^3 \left({1\over\lambda_j^2(t)}-
{1\over \lambda_1(t)\lambda_2(t)\lambda_3(t)}\right) \partial_{r_j}^2
\Phi(\vec{r},0).
\end{equation}
In the Thomas-Fermi approximation the spatial derivatives of $\Phi(\vec{r},0)$ are
neglected and the source
term vanishes; in this case, $\delta\tilde{\Phi}(\vec{r},t)$ being initially
zero remains zero, and $\tilde\Phi(\vec{r},t)$ remains constant\cite{STAB}: 
\begin{equation}
\tilde{\Phi}(\vec{r},t) \simeq \Phi(\vec{r},0).
\end{equation}
This result provides a generalization of the Thomas-Fermi approximation
to time dependent problems.
All the dynamics of the macroscopic wave function is contained 
in the evolution of three scaling parameters. In particular the
condensate density is a time dependent inverted paraboloid:
\begin{equation}
\label{TFT}
N|\Phi(\vec{r},t)|^2_{TF} = {\mu-\sum_{j=1}^3 {1\over 2}m\omega_j^2(0)
r_j^2/\lambda_j^2(t)\over g\lambda_1(t)\lambda_2(t)\lambda_3(t)}
\end{equation}
when the right hand side is positive and $|\Phi|^2_{TF}=0$ otherwise.

We now apply the above results to experimental data obtained 
in the Ioffe-Pritchard
trap of \cite{BEC4}. The trap is axially symmetric with respect to $z$
and cigar-shaped ($\omega_1=
\omega_2\equiv \omega_\perp\gg \omega_3 \equiv\omega_z$).

We consider first the simplest case of a sudden and total opening of the trap
at $t=0$. The equations for the evolution of the scaling parameters
(\ref{lambda}) simplify to:
\begin{eqnarray}
\label{cigarlambda}
{d^2 \over d \tau^2} \lambda_\perp &=&
{ 1 \over \lambda_\perp^3 \lambda_z } \nonumber\\
{d^2 \over d \tau^2} \lambda_z& =&
{\epsilon^2 \over \lambda_\perp^2 \lambda_z^2}
\end{eqnarray}
where $\lambda_{\perp}$ stands for $\lambda_1=\lambda_2$ and
$\lambda_z$ stands for $\lambda_3$. We have introduced a dimensionless time
variable $\tau=\omega_{\perp}(0)t$ and a parameter 
$\epsilon=\omega_z(0)/\omega_\perp(0)\ll 1$.
We solve (\ref{cigarlambda}) by an expansion in powers of $\epsilon$.
To zeroth order 
in $\epsilon$, $\lambda_z=1$ and the radial expansion scales as
\begin{equation}
\label{perp}
\lambda_\perp(\tau)=\sqrt{1 +\tau^2}.
\end{equation}
To second order in $\epsilon$ 
the axial expansion of the cloud is given by
\begin{equation}
\label{para}
\lambda_z(\tau) =1+\epsilon^2 \left[\tau \arctan \tau -\ln \sqrt{1+\tau^2}\right]
+O(\epsilon^4).
\end{equation}
For the experiments considered the term in $\epsilon^2$ is
not negligible.

We have performed a fit of the images obtained
for two different times of flight in \cite{BEC4}.
We used an inverted paraboloid for the density
of the cloud, having as free parameters 
the radial width $W_\perp$, the axial width $W_z$
and the number of condensed atoms $N$. 
Fig.1 shows a cut along
the $x$ axis (that is at $z=0$) of 
the spatial density of the cloud integrated along $y$.
According to Eq.(\ref{TFT}) the aspect ratio of the cloud
is given by
\begin{equation}
{W_z(t)\over W_\perp(t)} = 
{\lambda_z(t)\sqrt{2\mu/m\omega_z^2(0)}
\over \lambda_\perp(t)\sqrt{2\mu/m\omega_\perp^2(0)}}
= {\lambda_z(t)\over \lambda_\perp(t)}{1\over \epsilon}.
\label{RATIO}
\end{equation}
The fit gives the values of this ratio for two different expansion times,
from which we calculate the two unknown 
frequencies $\omega_\perp(0)$ and $\omega_z(0)$,
using Eq.(\ref{perp}),(\ref{para}).
From $W_\perp$ we 
calculate $\mu$; the relation (\ref{mu}) then leads to
a scattering length of $a=42 \pm 15$ Bohr for sodium, in agreement with
earlier measurements \cite{BILL}.

\begin{figure}[htb]
\centerline{\includegraphics[trim={0 0 0 9cm},width=8.5cm,clip=]{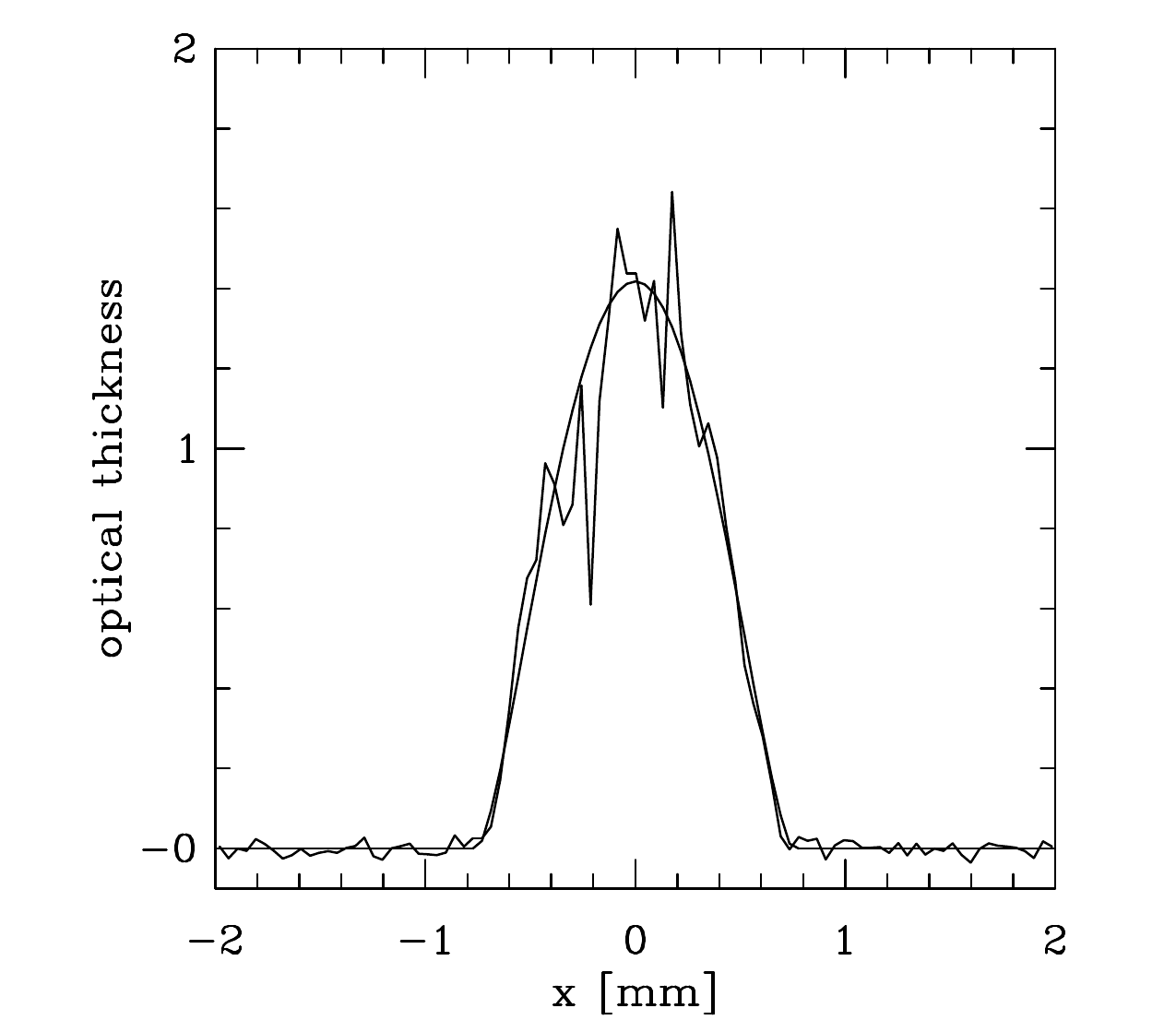}}
\caption{Spatial density of an expanding condensate integrated along $y$ axis,
cut along $x$ axis (that is at $z=0$). 
Experimental data obtained at MIT (expansion time of 40 ms) and
fit from theory.}
\end{figure}

In a second generation of experiments
a collective excitation of the condensate has been induced
by a time modulation of the eigenfrequencies of the trapping potential
\cite{JILA,MIT}. In \cite{MIT} the axial frequency is modulated as
$\omega_z^2(\tau)=\omega_z^2(0)[1-\eta(1-\cos\Omega \tau)]$ 
($\Omega$ in units of $\omega_\perp(0)$).
After the excitation the cloud freely oscillates in the unperturbed
potential for an adjustable time.
Finally, the trapping potential
is switched off and the expanding cloud is monitored.

By including this experimental sequence 
in the evolution of the scaling parameters (\ref{lambda}), we
give an {\it ab initio} calculation of the 
time of flight signals.
For a weak modulation 
($\eta\ll 1$) the 
time evolution in the trap is obtained
from a linearization of Eq.(\ref{lambda}) 
around the steady state value 1. During the excitation process we obtain 
for the deviations
$\delta\lambda_\perp$ and $\delta\lambda_z$:
\begin{eqnarray}
{d^2\over d\tau^2} \delta\lambda_\perp(\tau) &=& -4\delta\lambda_\perp(\tau)-
\delta\lambda_z(\tau) \label{FAST}\\
{d^2\over d\tau^2} \delta\lambda_z(\tau) &=& -2\epsilon^2
\delta\lambda_\perp(\tau)-3\epsilon^2\delta\lambda_z(\tau)\nonumber\\
& & + \epsilon^2\eta(1-\cos(\Omega\tau)).
\label{SLOW}\end{eqnarray}
To leading order in $\epsilon$ the eigenmodes of this linear system have 
frequencies (in units of $\omega_\perp(0)$)
$\Omega_{\mbox{\scriptsize fast}}=2$ and
$\Omega_{\mbox{\scriptsize slow}}
=\sqrt{5\over 2}\epsilon$
and are polarized along $(1,1,0)$ and $(1,1,-4)$ respectively
\cite{STRING,MODE}.
In the experiments the driving frequency $\Omega$ 
was set to $\Omega_{\mbox{\scriptsize slow}}$
to achieve a resonant excitation of the slow mode.
To lowest order in $\epsilon$ this allows to keep
only the slow mode component of the solution, for
which $\delta\lambda_\perp=-\delta\lambda_z/4$.
Eq.(\ref{SLOW}) integrated for the time duration $\tau_e$ of the excitation
then leads to:
\begin{equation}
\delta\lambda_z(\tau_e)={2\eta\over 5}(1-\cos(\Omega_{\mbox{\scriptsize slow}}\tau_e))
-{\eta\over 5}\Omega_{\mbox{\scriptsize slow}}
\tau_e\sin(\Omega_{\mbox{\scriptsize slow}}\tau_e).
\end{equation}
In \cite{MIT} the potential was modulated for five cycles so that
$\tau_e=5(2\pi/\Omega_{\mbox{\scriptsize slow}})$ 
and $\delta\lambda_z(\tau_e)=
0,{d\over d\tau}\delta\lambda_z(\tau_e)=
-2\pi\eta\Omega_{\mbox{\scriptsize slow}}$.
The subsequent evolution in the unperturbed trapping potential
is sinusoidal with the eigenfrequency $\Omega_{\mbox{\scriptsize slow}}$;
after the free oscillation time $\tau_o$ 
it leads to 
$\delta\lambda_z(\tau_e+\tau_o)=-4\delta\lambda_\perp(\tau_e+\tau_o)=
-2\pi\eta\sin\Omega_{\mbox{\scriptsize slow}}\tau_o$.

Finally the trapping potential is switched off 
to monitor the excited condensate.
The time evolution of the scaling parameters is obtained 
by linearizing Eq.(\ref{cigarlambda})
around the solutions Eq.(\ref{perp},\ref{para}) with 
initial conditions given by the values of $\delta\lambda_{z,\perp}$
and their derivatives at $\tau=\tau_e+\tau_o$.
After a time of flight
$\tau_f$ we obtain,
neglecting terms of order
2 in $\epsilon$ and terms vanishing in the limit $\tau_f\rightarrow
\infty$:
\begin{eqnarray}
\delta\lambda_\perp(\tau_e+\tau_o+\tau_f) &\hspace{-1.5mm}=\hspace{-1.5mm}&
-{1\over 4}\tau_f\ \delta\lambda_z(\tau_e+\tau_o)\nonumber\\
-{1\over 4}&\hspace{-1.5mm}[\hspace{-1.5mm}&\pi\tau_f-4\log\tau_f+1] {d\over d\tau}\delta\lambda_z(\tau_e+\tau_o)\nonumber\\
&\hspace{-1.5mm}\hspace{-1.5mm}& \\
\delta\lambda_z(\tau_e+\tau_o+\tau_f) &\hspace{-1.5mm}=\hspace{-1.5mm}& \delta\lambda_z(\tau_e+\tau_o)
+\tau_f {d\over d\tau}\delta\lambda_z(\tau_e+\tau_o) \nonumber\\
&\hspace{-1.5mm}\hspace{-1.5mm}& 
\end{eqnarray}
From this we determine the aspect ratio (\ref{RATIO})
of the expanding cloud.
Fig.2 shows that our predictions
are in good agreement with the experimental results
of \cite{MIT} for short free oscillation times $\tau_o$. 
For longer times $\tau_o$ 
a damping of the oscillations is observed experimentally, which
cannot be explained with our mean field treatment.

\begin{figure}[htb]
\centerline{\includegraphics[trim={0 0 0 8cm},width=8.5cm,clip=]{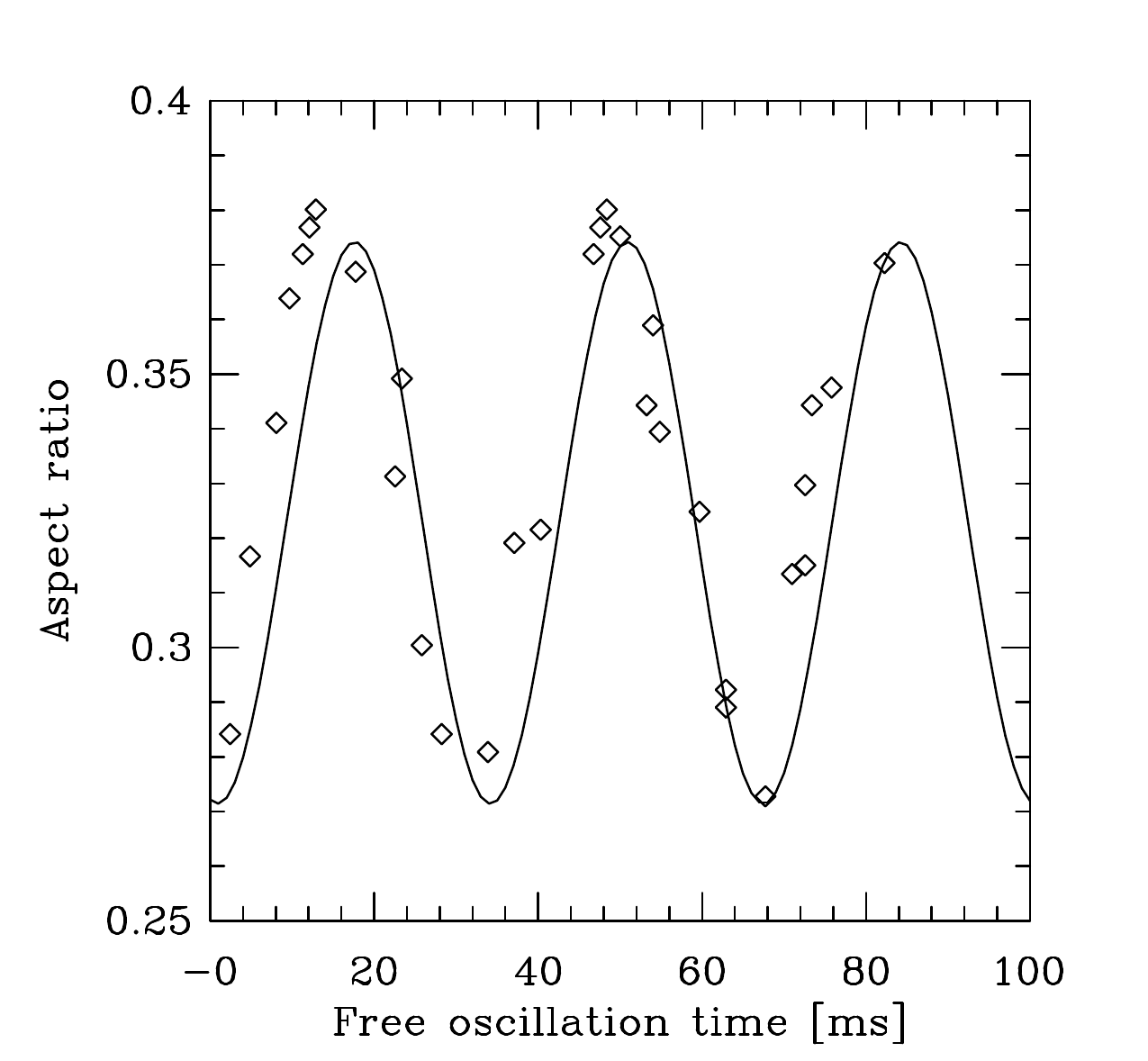}}
\caption{Aspect ratio of the excited and expanded condensate 
as a function of free oscillation time $\tau_o$.
Expansion time $\tau_f=40$ ms, 
$\omega_\perp(0)=2\pi\times 250$ Hz, $\omega_z(0)=2\pi 
\times 19$ Hz, $\eta=0.005$. Solid line: theory.
Diamonds: experimental
data obtained at MIT.}
\end{figure}

In conclusion, we have been able to extend the Thomas-Fermi
approximation to the motion of a condensate in a time dependent harmonic
potential: the time dependence is entirely contained in three scaling factors
which can be obtained from the evolution of a classical gas. 
This provides an easy quantitative tool for the analysis of current experiments
on trapped condensed gases. We have applied it to two recent experiments.
From time-of-flight images of the condensate at two different expansion times
we could calculate the scattering length $a$
without relying on 
measurements of the trapping frequencies.
For collective excitations of a condensate in the linear response regime 
we could predict not only the frequency but also the phase
and amplitude of the measured signal (see Fig.2).
Our treatment can be applied in the non-linear response regime as well,
for example for oscillations of the condensate induced by a strong
modulation or by a partial opening of the trap.

We are grateful to the group at MIT for providing us with the
experimental data. R.D.\ was supported by the European Community.
We thank M.-O.\ Mewes, J.\ Dali\-bard and C.\ Cohen-Tannoudji
for helpful discussions.


\begin{thebibliography}{99}
\bibitem{STAB} Linearizing the equation for $\delta\tilde{\Phi}(\vec{r},t)$ 
and assuming linear stability we obtain an upper bound
for $||\delta\tilde{\Phi}(t)||$ scaling in the isotropic case
as $\int_0^t dt'\,||S(t')||
\sim \int_0^t \omega(0) dt'|\lambda^{-2}(t')-\lambda^{-3}(t')| 
(\hbar\omega(0)/\mu)^{1/3}\rightarrow 0$ for $\mu/\hbar\omega(0)\rightarrow
\infty$.
\bibitem{MODE}
A third, axially
antisymmetric mode along $(1,-1,0)$ with a frequency $\sqrt{2}\omega_\perp$ 
is obtained from linearization of Eq.(\ref{lambda}) without assuming
$\lambda_1=\lambda_2$.
\end{thebibliography}
\end{document}